\documentclass[%
pre,
twocolumn,
superscriptaddress,
showpacs,
amsmath,amssymb,
aps,
showkeys,
]{revtex4-1}

\usepackage{graphicx}
\usepackage{dcolumn}
\usepackage{bm}
\usepackage{todonotes}
\usepackage{multirow}
\usepackage{subfigure}

\definecolor{pink}{RGB}{219, 48, 122}
\definecolor{purple}{RGB}{128,0,128}

\newcommand{\vect}[1]{\mbox{\boldmath $#1$}}

\begin{document}

\title{Graph spectral characterisation of the XY model on complex networks}

\author{Paul Expert}
\email{paul.expert08@imperial.ac.uk}
\thanks{These two authors contributed equally to this work.}
\affiliation{Department of Mathematics, Imperial College London, London, SW7 2AZ, UK}
\affiliation{EPSRC Centre for Mathematics of Precision Healthcare, Imperial College London, London, SW7 2AZ, UK}
\affiliation{Centre for Neuroimaging Sciences, Institute of Psychiatry, King's College London, London SE5 8AF, UK}
\author{Sarah de Nigris}
\email{denigris.sarah@gmail.com}
\thanks{These two authors contributed equally to this work.}
\affiliation{NaXys, D\'{e}partement de Math\'{e}matique, Universit\'{e} de Namur, 8 rempart de la Vierge, 5000 Namur, Belgium}
\affiliation{Univ Lyon, Cnrs, ENS de Lyon, Inria, UCB Lyon 1, LIP UMR 5668, F69342, Lyon, FRANCE}
\author{Taro Takaguchi}
\affiliation{National Institute of Information and Communications Technology, Tokyo 184-8795, Japan} 
\affiliation{National Institute of Informatics, Tokyo 101-8430, Japan}
\affiliation{JST, ERATO, Kawarabayashi Large Graph Project, Tokyo 101-8430, Japan}
\author{Renaud Lambiotte}
\affiliation{NaXys, D\'{e}partement de Math\'{e}matique, Universit\'{e} de Namur, 8 rempart de la Vierge, 5000 Namur, Belgium}

\date{\today}

\begin{abstract}
There is recent evidence that the XY spin model on complex networks can display three different macroscopic states in response to the topology of the network underpinning the interactions of the spins. In this work, we present a novel way to characterise the macroscopic states of the XY spin model based on the spectral decomposition of time series using topological information about the underlying networks. We use three different classes of networks to generate time series of the spins for the three possible macroscopic states. We then use the temporal Graph Signal Transform technique to decompose the time series of the spins on the eigenbasis of the Laplacian. From this decomposition, we produce spatial power spectra, which summarise the activation of structural modes by the non-linear dynamics, and thus coherent patterns of activity of the spins. These signatures of the macroscopic states are independent of the underlying network class and can thus be used as robust signatures for the macroscopic states. This work opens new avenues to analyse and characterise dynamics on complex networks using temporal Graph Signal Analysis.
\end{abstract}


\maketitle

\section{\label{sec:intro} Introduction}
Activity of brain regions \cite{Bullmore:2009iv}, car flow on roads \cite{Petri:2013bz}, meta-population epidemic \cite{Apolloni:2014jl}, all these arguably very different systems have in common that they can be represented as the activity of a quantity of interest on the nodes of a network. The coupling between the dynamics on the nodes and their network of interactions often leads to emergent collective states. In simple cases, such as the Kuramoto and $XY$ spin models, these macroscopic states can be classified according to the behaviour of an order parameter that measures the global coherence of the units comprising the system.

Unfortunately, this order parameter is blind to the structure of the underlying interaction network, and does not allow to investigate how the system behaves at different structural scales. To gain a better understanding of the effect of network structure on its activity, we need a method to characterise a macroscopic state that combines both aspects. Having such a method will then help us to understand the functioning and mitigate disruption of complex systems or even engineer new ones.

In this paper, we consider the problem in the spectral domain by exploiting tools from temporal Graph Signal Analysis. In particular, we show that the collective patterns of a dynamical system can be robustly characterised by decomposing its nodal activity in an adequate basis associated to its \emph{structural properties}.
 
To illustrate our method, we will study the $XY$ spin model on complex networks. Spin models are paradigmatic examples of systems where pairwise interactions give rise to emergent, macroscopic stationary states. Historically, the behaviour of these models was studied on lattices \cite{book_condensed_matter}, but other types of topologies have been considered in recent years and, in particular, researchers have investigated the effect of the topology on equilibrium and out of equilibrium states, for example, the Ising model \cite{exactsolIsing_lopes2004,herrero2002ising,Bianconi2002,Pekalski2011,Dorogotsev2002}, and the $XY$ model \cite{Medvedyeva:2003hh,Kim:2001jt,deNigrisPrE_2013,Kwak:2007jma}. Remarkably, a new stationary state has been observed on complex networks in addition to the well-known non-magnetised and magnetised states in the $XY$ spin model: the supra-oscillating state, in which magnetisation coherently oscillates indefinitely  \cite{deNigrisPrE_2013,deNigris2013EPL,de2014dimension,Kim:2001jt,Medvedyeva:2003hh,Kwak:2007jma,virkar2015hamiltonian,restrepo2014onset,berganza2013critical}. Interestingly, these three states can be found on several network models, which suggests that the topology constrains the $XY$ spins in a given phase.

Although it is possible to analytically connect the parameters of the network models to thermodynamics in some simple situations \citep{deNigrisPrE_2013}, the nonlinearity of the interactions in the $XY$ model complicates the construction of a direct, general theory linking the underlying network topology and the phenomenology. It is therefore desirable to develop a framework that links the structure to the dynamics.

We are in a situation where different topologies give rise to the same macroscopic states, and thus in a perfect setting to explore the interplay between structure and dynamics, using a network theoretical approach. As we are studying a global emergent property of a system, it is natural to seek a description that uses system-wide features of the underlying network. To explore the relationship between the structure of the network and the evolution of the individual spins, we leverage the spectral properties of networks and use the temporal Graph Signal Transform (tGST). Temporal Graph Signal Transform is an extension to time series of the Graph Signal Transform for static signals on complex networks \cite{ShumanIEEE2013,mcgraw2007analysis,mcgraw2008laplacian,asslani2012stochastic,asslani2013noise}. We use tGST to decompose the time series of the spins in the Laplacian eigenbasis, which carries information about the structure of the network.

The projection of the dynamics on the Laplacian eigenvectors has emerged in the realm of dynamical systems to successfully uncover, in reaction-diffusion and synchronization systems, complex patterns of activity on networks \cite{othmer1971instability,othmer1974non, mcgraw2007analysis,mcgraw2008laplacian,nakao2010turing,asslani2012stochastic,asslani2013noise,asllani2014theory,contemori2016multiple,nakao2014complex,cencetti2017topological,DiPatti2017}. On the other hand, in the signal processing community, the Graph Signal Transform has recently acquired resonance in the wider context of Graph Signal Processing with a more pronounced data driven streak. Indeed, in this latter field, such tools have already been applied, in different forms, for signal analysis such as fMRI time series \cite{behjat2014canonical} or image compression \cite{narang2012graph} as well as graph characterization and community detection \cite{tremblay2016subgraph,tremblay2014graph}. The crucial feature of the tGST, which explains its power and versatility, is its ability to analyse data on irregular domains such as complex networks.

By using tGST, we can quantify the importance of each eigenmode by computing the spatial power spectrum. We find that irrespective of the specific topology, the functional form of the power spectrum characterises a state. This clearly shows that a selection of modes is at play. In this paper, we will show  that the $XY$ dynamics resonates with specific graph substructures, leading to the same macroscopic state.

This paper is structured as follows: we briefly introduce the $XY$ spin model in Sec.~\ref{subsec:model} along with the several macroscopic behaviours it displays on networks in Sec.~\ref{subsec:phenomena}; we then proceed by introducing the general framework of the Graph Signal Transform in Sec.~\ref{subsec:GST} and finally present and discuss our findings in Sec.~\ref{sec:results}.

\section{Methods}\label{sec:methods}
\subsection{$XY$ spin model on networks}\label{subsec:model} 
We consider the $XY$ spin model, a well known model in statistical mechanics, on various network topologies. In this model, the dynamics of the spins is parametrised by an angle $\theta_{i}(t)$ and its canonically associated
momentum $p_{i}(t)$.
Each spin $i$ is then located on a network vertex
and interacts with the spins in $V_{i}$, the set of vertices
connected to $i$. The Hamiltonian of the system reads: 
\begin{equation}
H=\sum_{i=1}^{N}\frac{p_{i}^{2}}{2}+\frac{J}{2\left\langle k\right\rangle }\sum_{i=1}^{N}\sum_{j\in V_{i}}(1-\cos(\theta_{i}-\theta_{j})),\label{eq:hamiltonian}
\end{equation}
where $\theta_{i}\in[0,2\pi]$, $J>0$ and $\left\langle k\right\rangle $ is the average degree of the network. The dynamics is given by the following Hamilton's equations:
\begin{align}
\begin{cases}
\dot{\theta_{i}} & =p_{i},\\
\dot{p}_{i} & =-\frac{J}{\left\langle k\right\rangle }\sum_{j\in V_{i}}\sin(\theta_{i}-\theta_{j}),
\end{cases}
\label{eq:dynamics}
\end{align}
As we are in the microcanonical ensemble, the energy $H$ (Eq.~\eqref{eq:hamiltonian}) is conserved along with the total momentum $P \equiv \sum_{i}p_{i}/N$ which itself is conserved because of the translational invariance of the system.

In order to determine the amount of coherence in the system, we define the order parameter  $M=\left|\mathrm{\mathbf{M}}\right|$, where the magnetization $\mathbf{M}$ is defined by
\begin{equation}
\mathrm{\mathbf{M}} \equiv \left(\frac{1}{N}\sum_{i} \cos\theta_{i}, \  \frac{1}{N}\sum_{i} \sin\theta_{i}\right).
\label{eq:magnetization}
\end{equation}
In the stationary state, it is possible to measure $\overline{M}$,
where the bar stands for the temporal mean. In the magnetised phase, all rotors point in the same direction and $\overline{M}\sim 1$, while in the non-magnetised phase there is no preferred direction for the rotors and $\overline{M}\sim 0$.

\subsection{Phenomenology on networks}\label{subsec:phenomena}
\begin{figure}
\includegraphics[width=4cm]{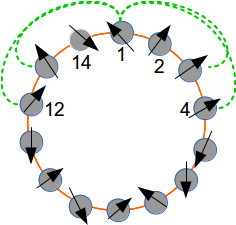}
\caption{Practical construction of a Lace Network for $N=14,k=2$ and $r=\lfloor\sqrt{N}\rfloor=3$. The starting configuration is in orange and the dotted green links are the possible rewirings allowed within the range $r$.}
\label{fig:lace}
\end{figure}

Our choice of the $XY$ model was motivated by the variety of macroscopic behaviours displayed when the rotors interact on a complex network at low energies. We now briefly recapitulate this phenomenology in order to give the background upon which the present work is based.
The behaviour of the $XY$ model on complex networks has been explored in Ref.~\cite{deNigrisPrE_2013,deNigris2013EPL,de2014dimension,belger2016slowing} where the authors considered the thermodynamics on three different topologies: $k$-regular networks, Watts-Strogatz (WS) small-world networks \cite{watts_strogatz1998SW}, and Lace networks~\cite{de2014dimension}. Here, a $k$-regular networks refers to a network where nodes are arranged on a one-dimensional ring and connected to their $(k/2)$ next nearest neighbours on each side \cite{deNigris2013EPL}. Lace networks are generated from a $k$-regular network where each link can be rewired with probability $p$ to a node within  a range $r\propto \lfloor N^\delta \rfloor,\, 0<\delta\leq1$ where the distance is measured by hops along the original ring (see Fig.~\ref{fig:lace}). The Lace network model is a variant of the WS network model with an additional constraint on the rewiring process. In the following we will review the behaviour of the $XY$ spin model on the three network models we considered, see Table~\ref{tab:topo_phases} for a summary of the topological conditions for each phase to exist.

\begin{figure}
	\includegraphics[width=\columnwidth]{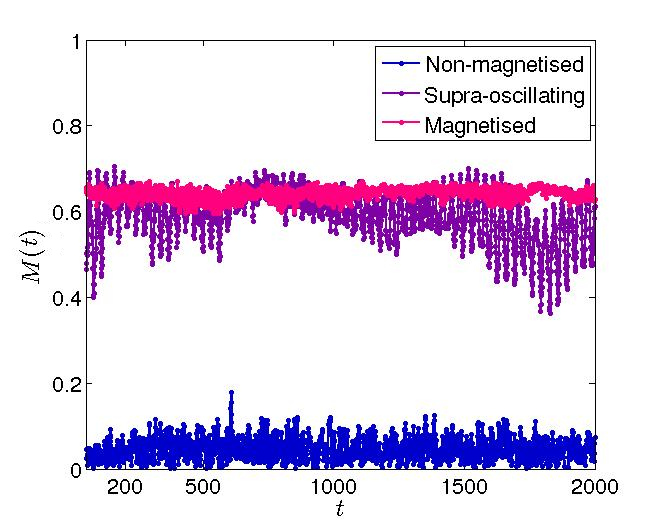}
	\caption{Temporal behavior of the magnetization $M(t)$ for the three asymptotically stable regimes of the $XY-$model: (bottom line) a non-magnetized, (middle line) supra-oscillating and (top line) magnetized. The underlying topology here is a Lace network of 16384 nodes  at $\varepsilon=0.365$ with parameters: non-magnetised $\delta=0.5,\ p=0.001$, supra-oscillating $\delta=0.5,\ p=0.9$ and magnetised $\delta=0.75,\ p=0.9$.}\label{fig:macro-states}
\end{figure}

\paragraph*{(i) $k$-regular graphs} The nodal degree $k\propto \lfloor N^\beta \rfloor $, with $0<\beta\leq1$, determines the stationary state of the model. A non-magnetised phase is present at all energies $H$ for $\beta < 0.5$. By contrast, a magnetised phase is observed for $\beta > 0.5$ and low energies densities, i.e. $\varepsilon=H/N \lesssim 0.8$~\cite{deNigris2013EPL}. For $\beta = 0.5$, a highly oscillating state emerges, in which the order parameter $M$ is affected by persistent macroscopic fluctuations (Fig.~\ref{fig:macro-states}) \cite{deNigris2013EPL,de2014dimension}. At odds with the classic behaviours, these fluctuations are persistent over time: they have been observed in simulations up to 10 times longer than the usual relaxation to equilibrium time. Moreover, it has been shown that they are not due to finite-size effects: the fluctuations cannot be tamed by increasing the system size as their variance displays a remarkable stability across different system sizes \cite{deNigrisPrE_2013,belger2016slowing}. We will refer to this state as supra-oscillating state.

\begin{figure}
\includegraphics[width=9.3cm,trim={3mm 0 0 0},clip]{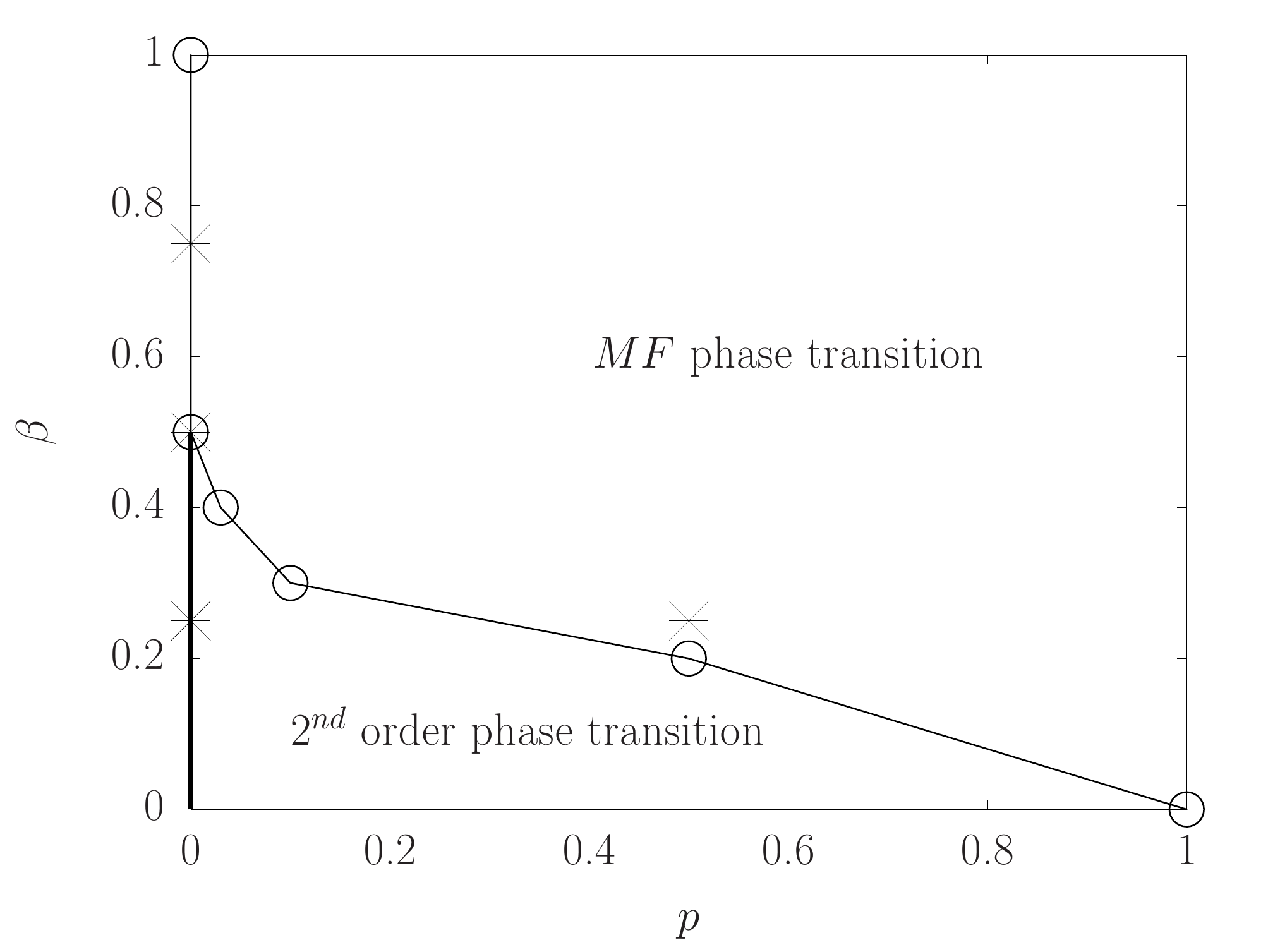}
\caption{Phase portrait of the $XY$-model on $k$-regular and Small World. On the $x$ axis the rewiring probability of the WS Small World model and on the $y$ axis the $\beta$ parameter giving the degree $k\propto N^{\beta}$. The bold line for $0<\beta<0.5$ indicates the non magnetised regime. The dot at $\beta=0.5$ represents the oscillating regime. The stars mark the parameters set of the $k$-regular and WS networks used in Figs.~\ref{fig:spectra_powa_across}-\ref{fig:spectra_powa_stat}. Finally, ``MF'' stands for the classical mean field second order phase transition, occurring at $\varepsilon=0.75$, while in the region marked by ``$2^{nd}$ order phase transition'' the transition energy is affected by the $(\beta,p)$ parameters \cite{deNigrisPrE_2013}.}
\label{fig:phase-portrait}
\end{figure}

\paragraph*{(ii) WS networks} The only topological parameter governing the thermodynamics is the rewiring probability $p$. When $p>p_{\rm WS}\propto 1/N^{\beta+1}$, the network possesses the small-world property. The proportion of long-range links introduced by the rewiring process increase the effective correlation length and the system is in a magnetised state for all values of $\beta$ (Fig.~\ref{fig:phase-portrait}). For $p<p_{\rm{WS}}$, we recover the existence of a non-magnetised phase. Interestingly, it is not possible to observe the supra-oscillating state on a WS network between $p<p_{\rm{WS}}$ and $p>p_{\rm{WS}}$ as in the case of the $k$-regular graph, for any value of $\beta$. Even a small amount of unconstrained randomness has a homogenizing effect on the dynamics, making the potential the parameter space in which the supra-oscillating state could live is extremely small and precludes its observation in practice.

\paragraph*{(iii) Lace networks:}Lace networks sit on the boundary between the $k$-regular graphs and WS networks. The constraint on the rewiring range partially preserves the regularity of the k-regular graph. This additional constraint enables the network to retrieve the supra-oscillating state that disappears for WS networks. It is worth stressing that Lace networks display the supra-oscillating state without the need for a density of links which was necessary for the $k$-regular networks. All the results for Lace networks are actually obtained in an extremely sparse setting, namely $\beta=0.2$. The reason is that for those networks the crucial parameter is the rewiring range $\delta$: for $\delta = 0.5$, a rewiring probability $p^{*}$ exists above which the supra-oscillating phase sets in. The non-magnetised and magnetised phases are separated by the same probability $p^{*}$ for $\delta > 0.5$, as long range interactions are needed to keep the system in the magnetised phase. In the $\delta<0.5$ case, the rewiring range is too short to allow the emergence of a coherent state. Finally, we note that the probability $p^*$ depends on the size of the network, as it is related to an effective ``network dimension'' \cite{de2014dimension}. It is therefore not possible to have a phase portrait for Lace networks as it is for $k$-regular and WS network in Fig.~\ref{fig:phase-portrait}. In the present study, the parameters $\delta$ and $p$ are chosen according to the system size $N$ in order to obtain a specific target state.

\begin{table}
\begin{center}
\caption{Summary of the topological conditions leading to a specific macroscopic state: for WS networks, $p_{WS}\propto 1/N^{\beta+1}$ \cite{deNigrisPrE_2013} and, for lace networks, $p^* (N,r)$ is the probability required for the network to display an effective dimension of 2 \cite{de2014dimension}.
For a given $N$, the exponents $\beta$ and $\delta$ determine $k$ and $r$ as $k \propto N^\beta$ and $r \propto N^\delta$, respectively. With the values listed in the table below, these conditions yield, for $N=2048$,  $k=45$ for the SO generated by $k$-regular networks and $r=45$ in the Lace generated SO state. For all Lace networks we use $\beta=0.2$ in order to obtain sparse networks. In the left column, M stands for magnetised, SO for supra-oscillating and NM for non magnetised.}

\begin{tabular}{ |c|c|c|c| } 
\hline
State & $k$-regular & WS & Lace\\
\hline
M& $\beta > 0.5$ & $p > p_{\rm WS}$ & $\delta > 0.5 \wedge p\geq p^*$\\ 
SO & $\beta = 0.5$ & ---  & $\delta = 0.5 \wedge p\geq p^*$\\
NM  & $\beta < 0.5$ & $\beta < 0.5\wedge p < p_{\rm WS}$ & $\delta < 0.5, \delta\geq 0.5 \wedge p<p*$\\ 
\hline
\end{tabular}\label{tab:topo_phases}
\end{center}
\end{table}

\subsection{Temporal Graph Signal Transform}\label{subsec:GST}
In the previous section, we summarised the topological conditions for each phase of the $XY$ spin model on the network models we considered. The three phases are identified by the behaviour of the magnetisation which is blind to the finer coherence patterns in the evolution of the spins. From the conditions on the parameters summarised in Table \ref{tab:topo_phases} to generate underlying networks, it is evident that topology plays a crucial role in the emergence of a specific phase. It is therefore natural to use structural information about the underlying network to characterise each phase. To disentangle the relationship between the structure and the macroscopic behaviour induced by the dynamics, we will use the temporal Graph Signal Transform (tGST) approach to highlight the importance of whole network structures to explain the temporal evolution of the orientation of the spins.

Before going into the specifics of the present analysis, we present the general framework of tGST. Let us suppose that we have a static, undirected graph $G$ in which the state of the nodes change over time. The activity of node~$i$ at time $t$ is denoted by the scalar variable $x_{i}(t)$ and the state of the system is represented by the vector $\vect{x}(t) \equiv \left( x_{1}(t), x_{2}(t), \ldots, x_{N}(t) \right)$. When the activities of the nodes are coupled, it is then convenient to use a $N \times N$-matrix $\cal{A}$, whose $(i,j)$-element describes the interaction between nodes $i$ and $j$, such that the evolution of $\vect{x}$ can be written in all generality as $\vect{x}(t+1)=F\left(\cal{A},\vect{x}\right)$. For instance, the $XY$ model dynamic equations in Eq.~\ref{eq:dynamics} can be recast in matrix form as
\begin{align}
\begin{cases}
\vect{\dot{\theta}} & =\bm{p}\,\\
\bm{\dot{p}} & =-\frac{J}{\left\langle k\right\rangle }\bm{\sin(\theta)}^\intercal\cal{A}\bm{\cos(\theta)}-\bm{\cos(\theta)}^\intercal\cal{A}\bm{\sin(\theta)}),
\end{cases}
\end{align}
where $\cal{A}$ is the underlying network adjacency matrix, $\bm{\sin(\bm{\theta})}=(\sin(\theta_{1}),\ldots,\sin(\theta_{N}))^\intercal$ and similarly for $\bm{\cos(\bm{\theta})}$ and $^\intercal$ denotes the vector transposition.
Let us furthermore suppose that the matrix $\cal{A}$ is real and symmetric. By the spectral theorem, $\cal{A}$ has the eigenvectors $\left\{ v_\alpha \right\}$ and their associated eigenvalues $\lambda_\alpha$ ($\alpha = 0,2, \cdots, N-1$). The Graph Signal Transform consists in projecting the activities $\vect{x}(t)$ at time $t$ onto the set of eigenvectors $\vect{v}_{\alpha}$:
\begin{equation}
\hat{x}_\alpha(t) = \sum_{i=1}^{N} x_{i}(t) v_{\alpha}^{i},
\label{eq:GST_x}
\end{equation}
where $v_{\alpha}^{i}$ is the $i$-th component of $\vect{v}_{\alpha}$.
Up to proper normalization of $\vect{x}(t)$, $\hat{x}_\alpha(t)$ can be interpreted as the similarity between the signal $\vect{x}(t)$ and the structure described by $v_\alpha$. Common examples of the matrix $\cal{A}$ are the adjacency matrix $A$, describing the connectivity of the network or the Laplacian matrix $L$, governing diffusion processes. The Laplacian $L$ is defined by $L \equiv D - A$, where is $D \equiv {\rm diag}(k_{1}, k_{2}, \ldots, k_{N})$ and $k_i$ is the degree of node $i$. 

The choice of interaction matrix, and its associated eigenbasis for tGST, can be chosen freely by the user: the choice of the operator used to decompose the signal, e.g. the adjacency matrix or a Laplacian type operator, will emphasis different aspect of the original signal due to their intrinsic filtering properties \cite{gavili2015shift}.
In this study, we used the Laplacian eigenbasis to decompose the time series of the spins in the three states of the XY-model, because from an operator point of view, it quantifies the \emph{signal smoothness} \cite{smola2003kernels}.
The Laplacian operator has been used to study systems close to synchronisation: for example, the dynamics of Kuramoto oscillators interacting on networks via diffusive coupling close to synchronisation can be linearised and written with the Laplacian $L$. The Laplacian eigenbasis is therefore a natural choice to investigate the synchronization phenomena \cite{Kuramoto1984,Almendral:2007bm,RodriguesPhysRep2016}. Examples of such analysis include the master stability function analysis of the synchronous states \cite{PecoraPRL1998,FinkPRE2000,BarahonaPRL2002}, the effects of network structure on synchrony \cite{McGrawPRE2007,McGrawPRE2008,KalloniatisPRE2010,ZuparicPhysD2013}, and the synchrony of non-identical oscillators \cite{FujiwaraEPJB2009}.

We also note that the $XY$ dynamics in the magnetised state can be well approximated by the Laplacian dynamics, but we stress that using the Laplacian operator to decompose the time series yields compelling results for all phases, including when the dynamics is non-linear, indicating that the tGST approach is versatile and a powerful and generic tool to analyse time series on networks.
 
The correspondence between the GST and the discrete spatial Fourier transform is evident from Eq.~\ref{eq:GST_x}. This gives the keys for a better grasp of signal smoothness and how eigenvectors are be used to represent signal at different levels of granularity. Indeed, each eigenvector $\vect{v}_{\alpha}$ represents a specific weighted node structure associated to the corresponding eigenvalue $\lambda_{\alpha}$. In this context, the eigenvalue can be interpreted as a coherence length, in the same way that each Fourier mode is associated to a wave of specific wavelength. With this parallel in mind, a small eigenvalue is associated to a long wavelength; the extreme example being the zero eigenvalue $\lambda_0=0$ that corresponds to the connectedness of the network and to the uniform eigenvector $v_0\propto(1\dots1$). Due to the possible degeneracy of the eigenvalues, it is in general not possible to establish a bijection between the ``wavelength'', i.e., the eigenvalue, and the ``wave'', i.e., the eigenvector. This is for example the case for $k$-regular graphs that have very high symmetries (see section~\ref{subsec:excitation}). We also note that as the eigenvalues of the Laplacian satisfy $0=\lambda_{1}\leq\lambda_{1}\leq\ldots\leq\lambda_{N}$, the eigenvalues are naturally ordered by decreasing ``wavelength''.

\section{Results}\label{sec:results}
In this section, we apply the temporal Graph Signal Transform (tGST) using the Laplacian eigenbasis to the time series generated by the $XY$ spin model in the three stationary phases on the three network models. By doing this analysis, we aim at: (i) finding the eigenmodes that support each macroscopic phases, (ii) investigating the spectral features of each phases that are commonly observed across different network models, and (iii) interpreting these spectral features in a geometric or structural way.

\subsection{Numerical simulations}\label{subsec:numerics}
In this section, we describe the specifics of the numerical simulations we performed. We run molecular dynamics (MD) simulations of the isolated system in Eqs.~(\ref{eq:dynamics}), starting with Gaussian initial conditions $\mathcal{N}(0,T)$ for $\left\{ \theta_{i},p_{i}\right\}$, with $T$ the temperature. With these initial conditions, the total momentum $P$ is therefore set at $0$ without loss of generality. The simulations are performed by integrating the dynamic equations in Eqs.~(\ref{eq:dynamics}) with the fifth order optimal symplectic integrator, described in \cite{Mclachlan92}, with a time step of $\Delta t=0.05$. This integrating scheme allows us to check the accuracy of the numerical integration; we verified at each time step that the conserved quantities of the system, energy $H$ and total momentum $P$ are effectively constant over time. Once the network topology and the size $N$ are fixed, we monitor the average magnetization $\overline{M}(\varepsilon)$ for each energy $\varepsilon=H/N$ in the physical range. We compute the temporal mean $\overline{M}$ on the second half of the simulation, after checking that the magnetization has reached a stationary state and only use this part of the simulations in our analyses. The simulation time $t_f$ is typically of order $O(10^{5})$. Finally, the system sizes considered for our analyses range from $N=2^{10}$ to $N=2^{13}$.

\subsection{Laplacian modes excitation}\label{subsec:excitation}
As detailed in Sec.~\ref{subsec:phenomena}, the $XY$ spin model possesses the potential to display three macroscopic regimes in response to different network topologies, and Fig.~\ref{fig:macro-states} demonstrates the typical temporal behaviour of the magnetization $M$ for the magnetized, non-magnetized and supra-oscillating phases. We point out that the fluctuations of the magnetization are due to finite size effects and only disappear in the thermodynamic limit, with the exception of the supra-oscillating state, for which they are intrinsic \cite{deNigrisPrE_2013,belger2016slowing}.

\begin{figure*}
	\includegraphics[width=0.3\textwidth]{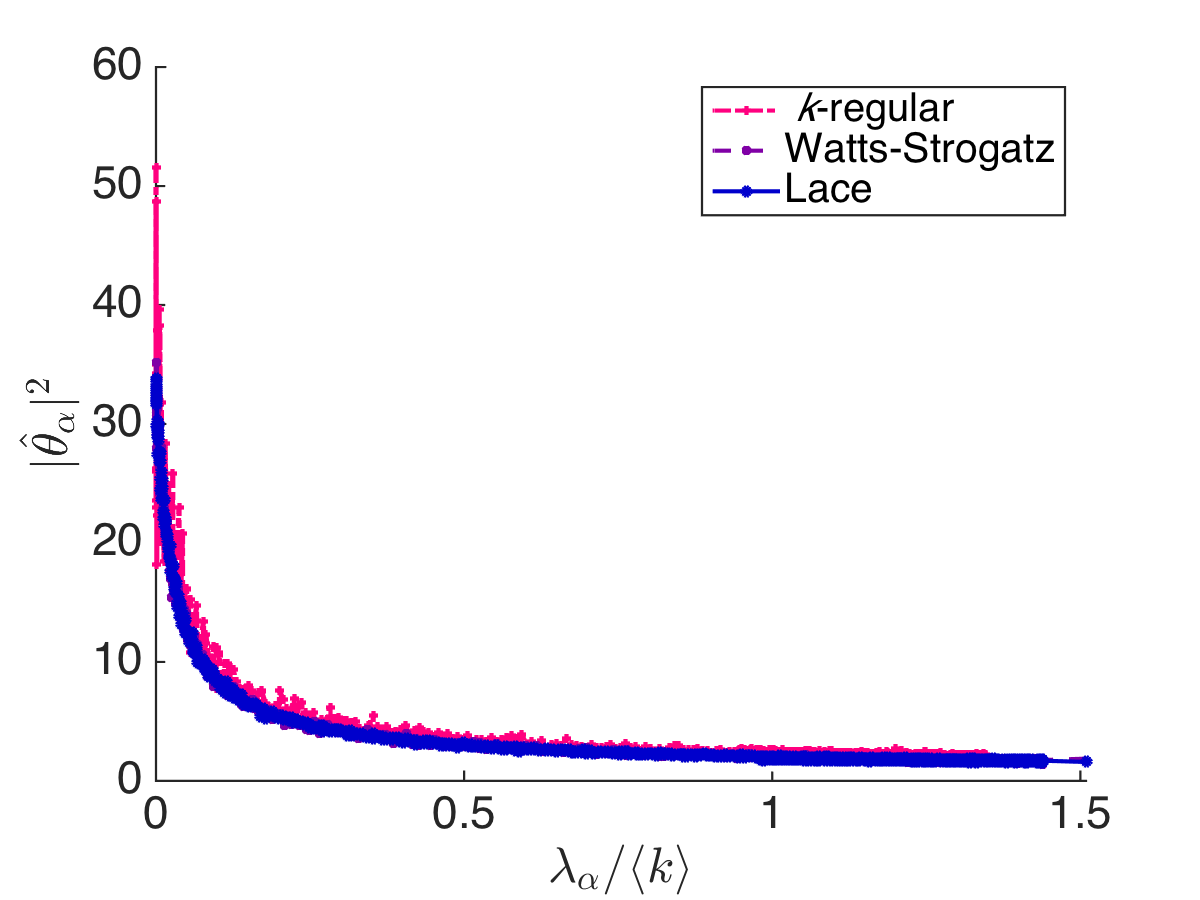}
	\includegraphics[width=0.3\textwidth]{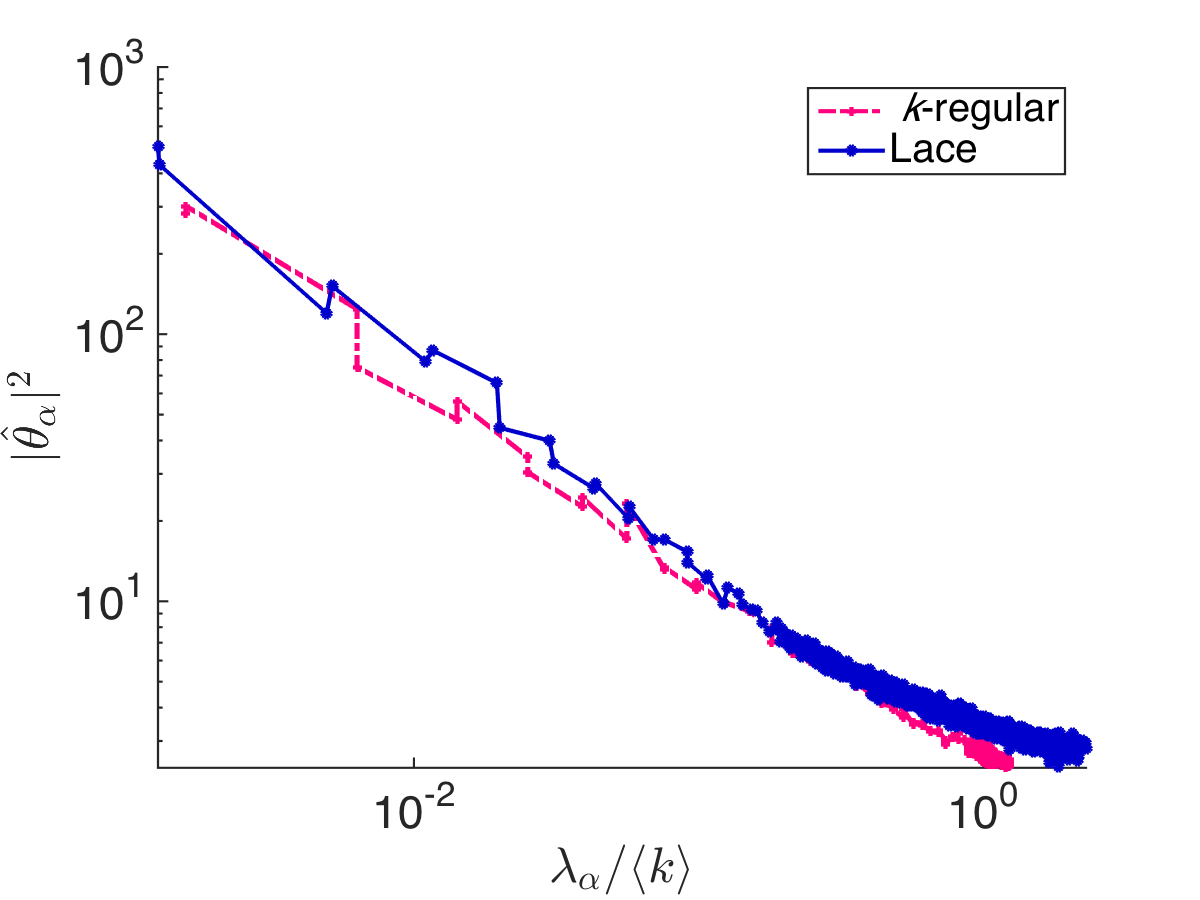}
	\includegraphics[width=0.3\textwidth]{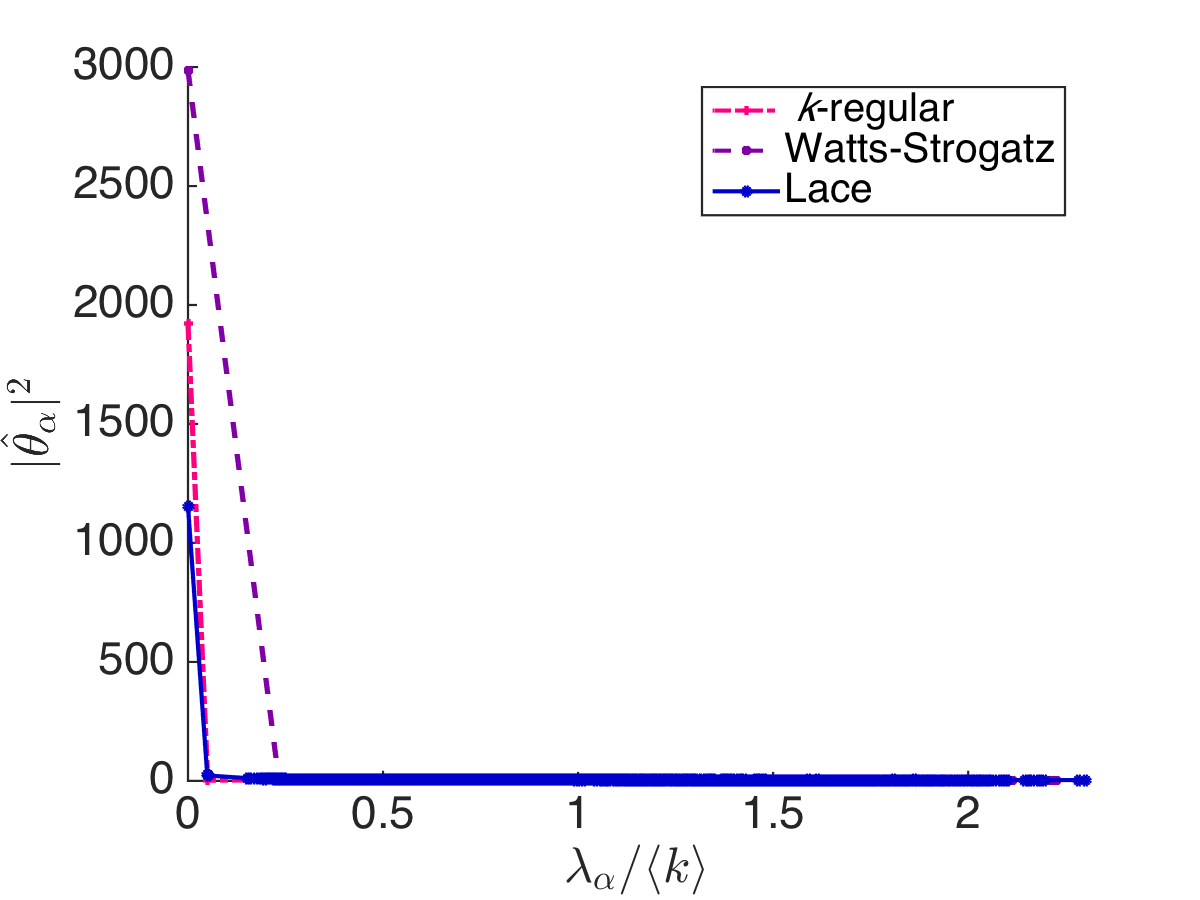}\\
	\vspace{.5cm}
	\includegraphics[width=0.3\textwidth]{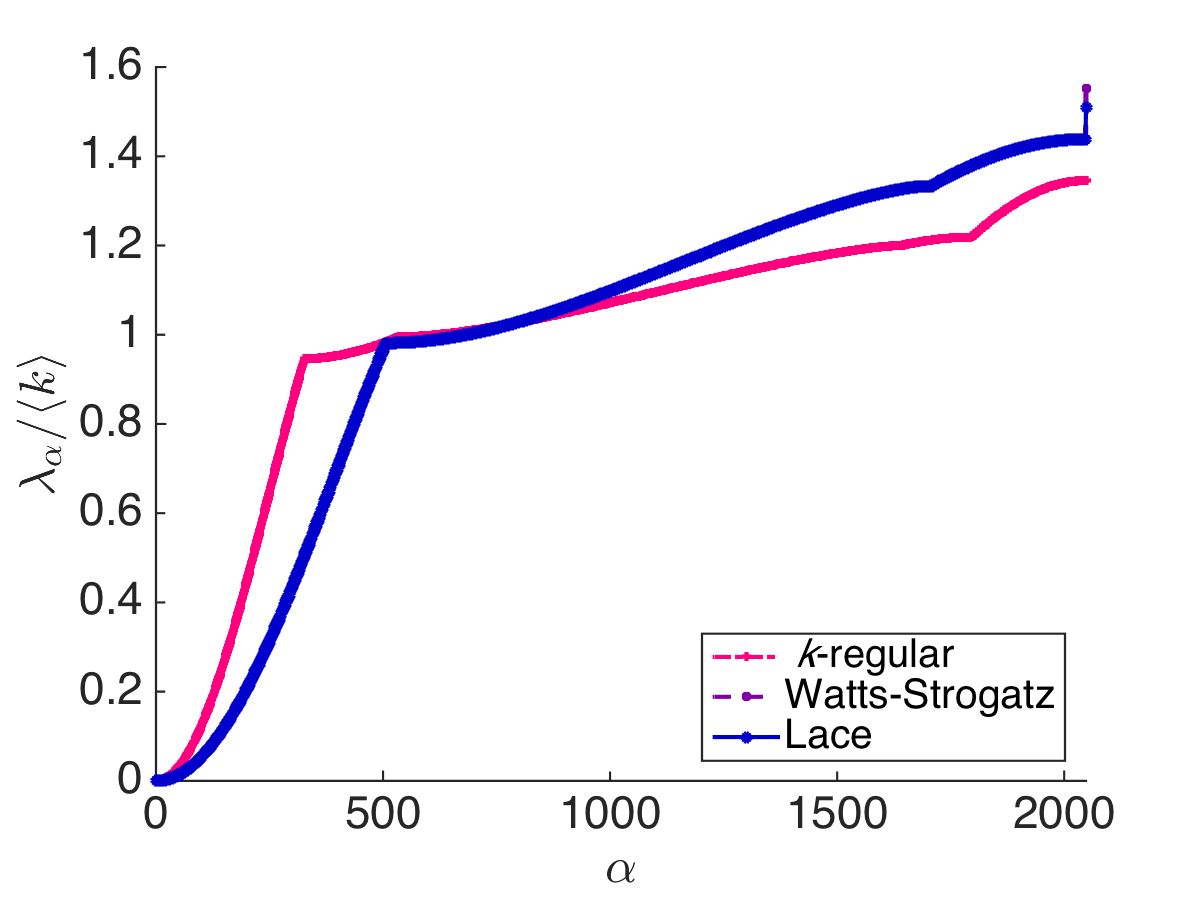}\put(-77,-12){a)}
	\includegraphics[width=0.3\textwidth]{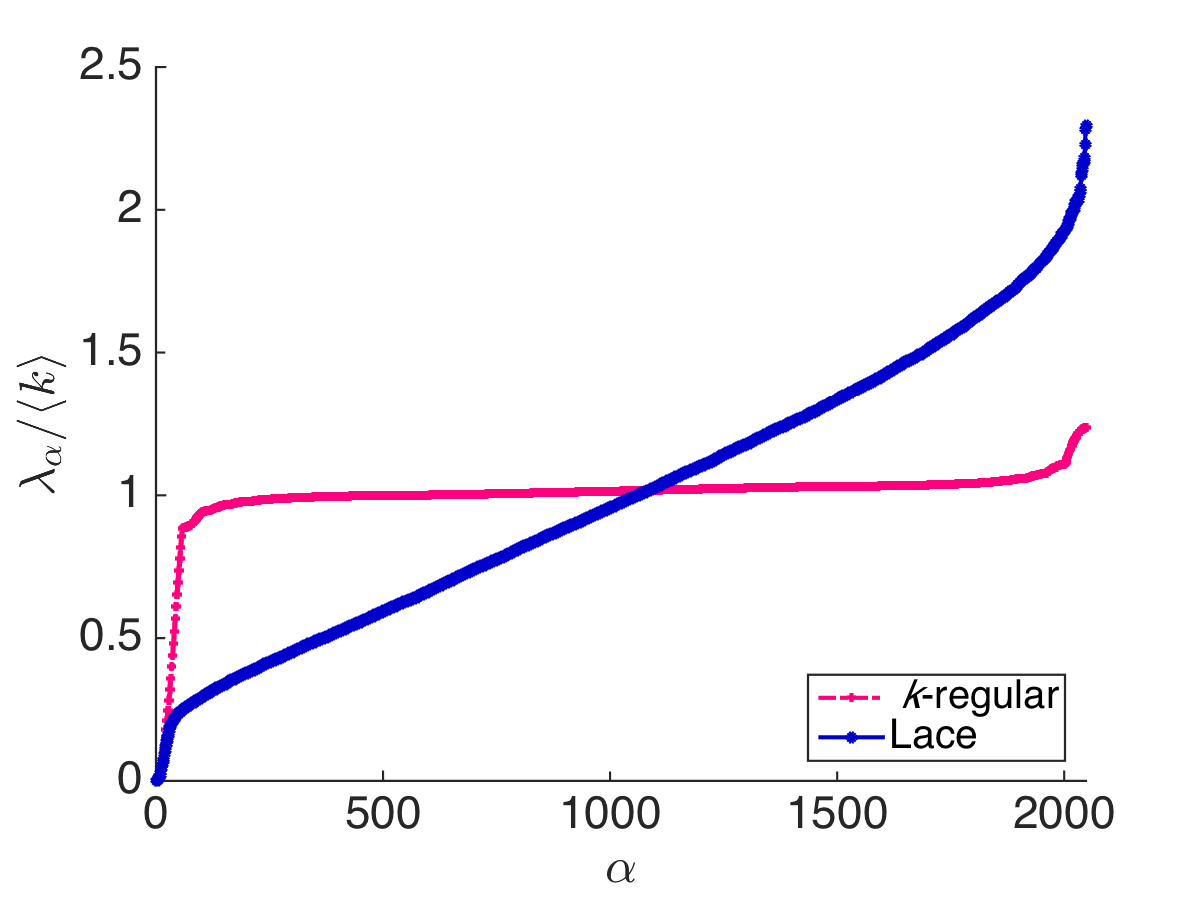}\put(-77,-12){b)}
	\includegraphics[width=0.3\textwidth]{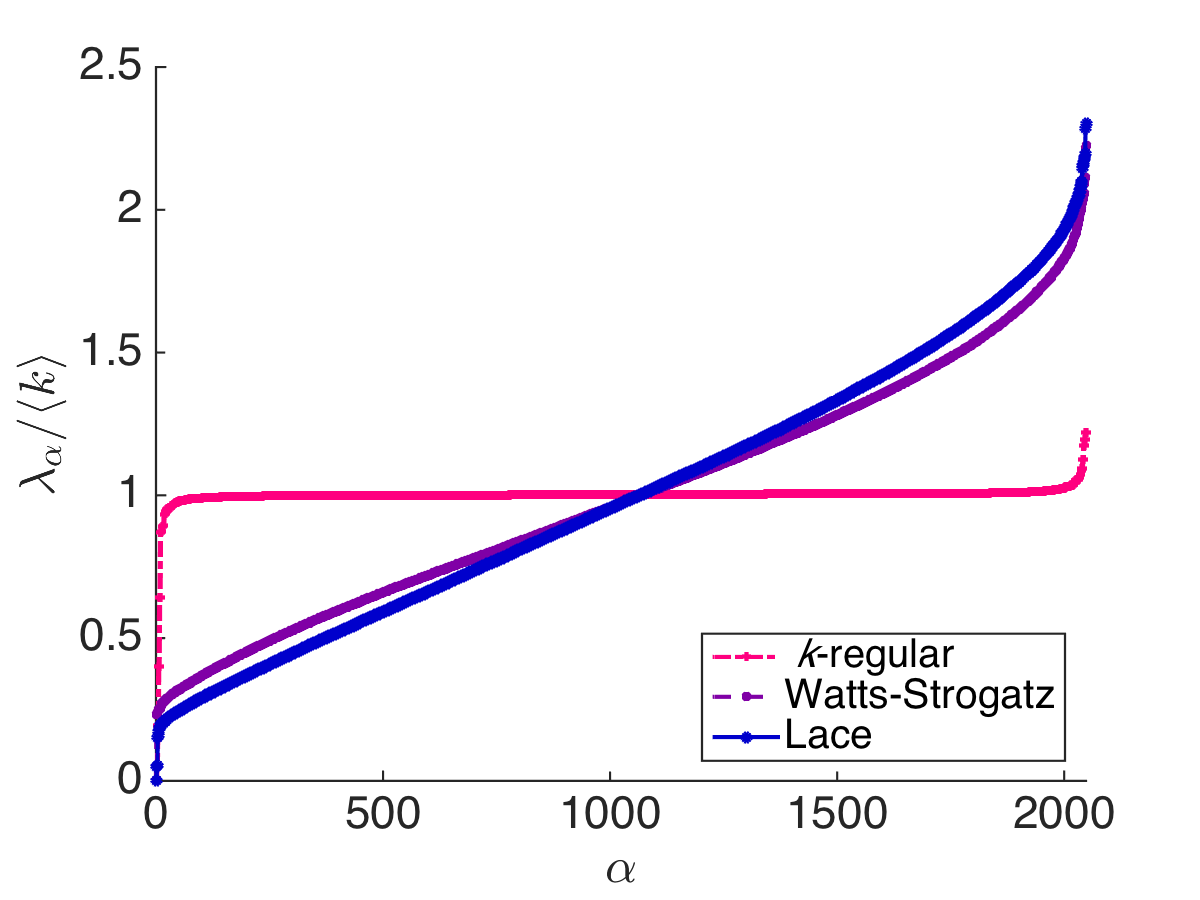}\put(-77,-12){c)}
	\caption{In this figure, we show the spatial power spectra (upper row) and the spectra (lower row) for the different macroscopic states and network topologies we consider in this article: $k$-regular, Watts-Strogatz and Lace networks. a) Non-magnetised state, the spectra for Lace and WS are quasi overlapping. b) Supra-oscillating state without Watts-Strogatz networks as they do not have parameters that accommodate this state. The power spectra are displayed in log-log scale to show the hierarchy in eigenmodes. c) Magnetised state. In this figure the networks have size $N=2048$ and topological parameters: Lace Networks $\beta=0.2$ a) $\delta=0.5$, $p=0.0001$,b)  $\delta=0.5$, $p=0.5$, c) $\delta=0.75$, $p=0.5$; $k$-regular a) $\beta=0.25$, b) $\beta=0.5$, c) $\beta=0.75$; WS a) $\beta=0.25$, $p=0.00007$ c) $\beta=0.25$, $p=0.5$.}\label{fig:spectra_powa_across}
\end{figure*}

\begin{figure*}
	\includegraphics[width=0.3\textwidth]{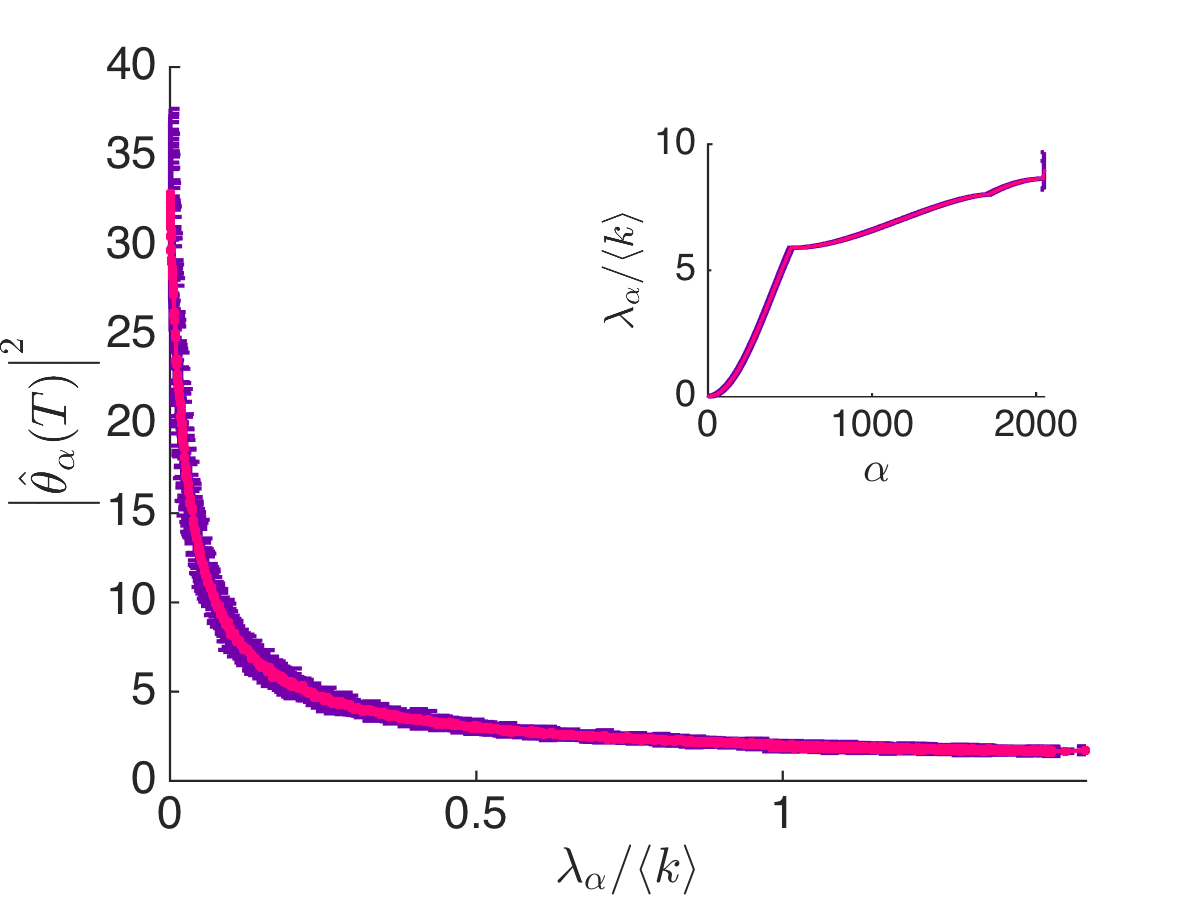}\put(-77,-12){a)}
	\includegraphics[width=0.3\textwidth]{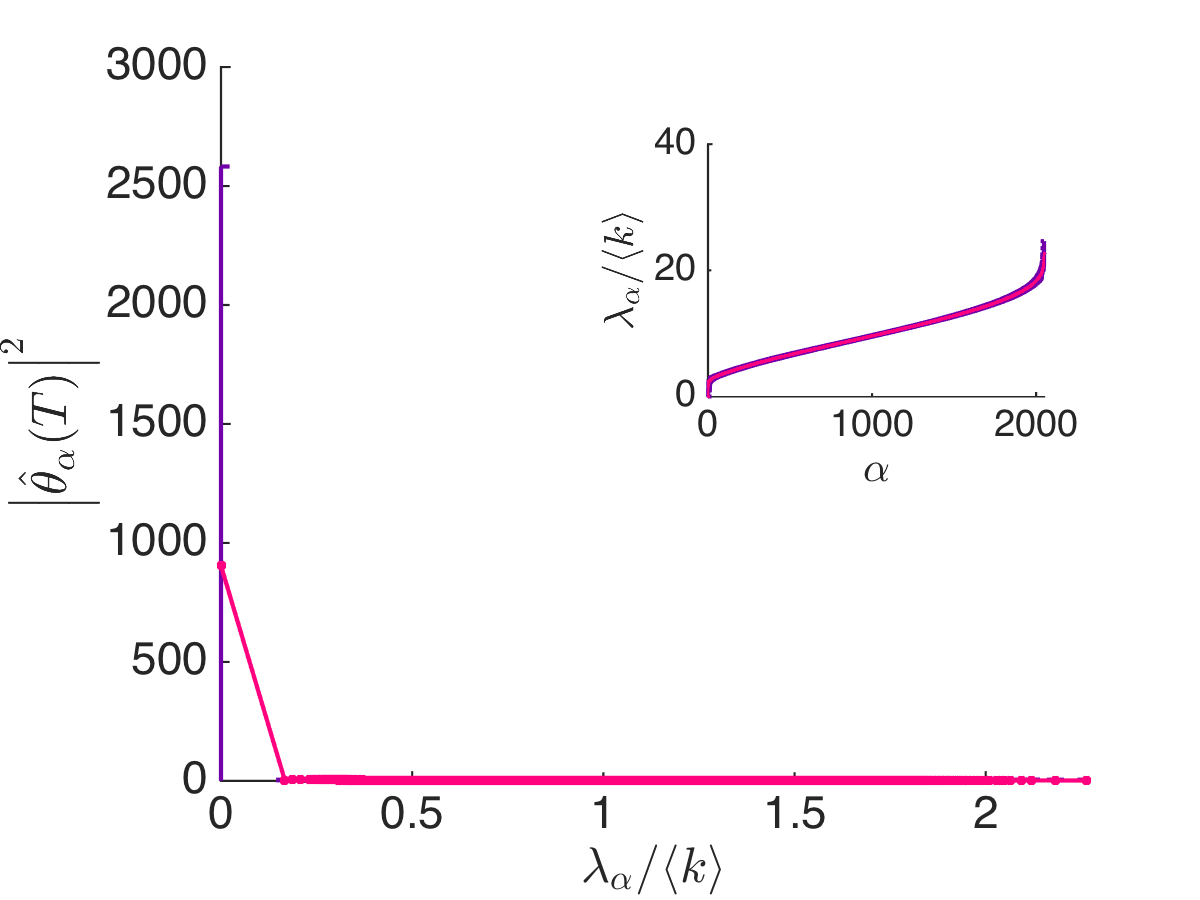}\put(-77,-12){b)}\\
    \vspace{.5cm}
	\includegraphics[width=0.3\textwidth]{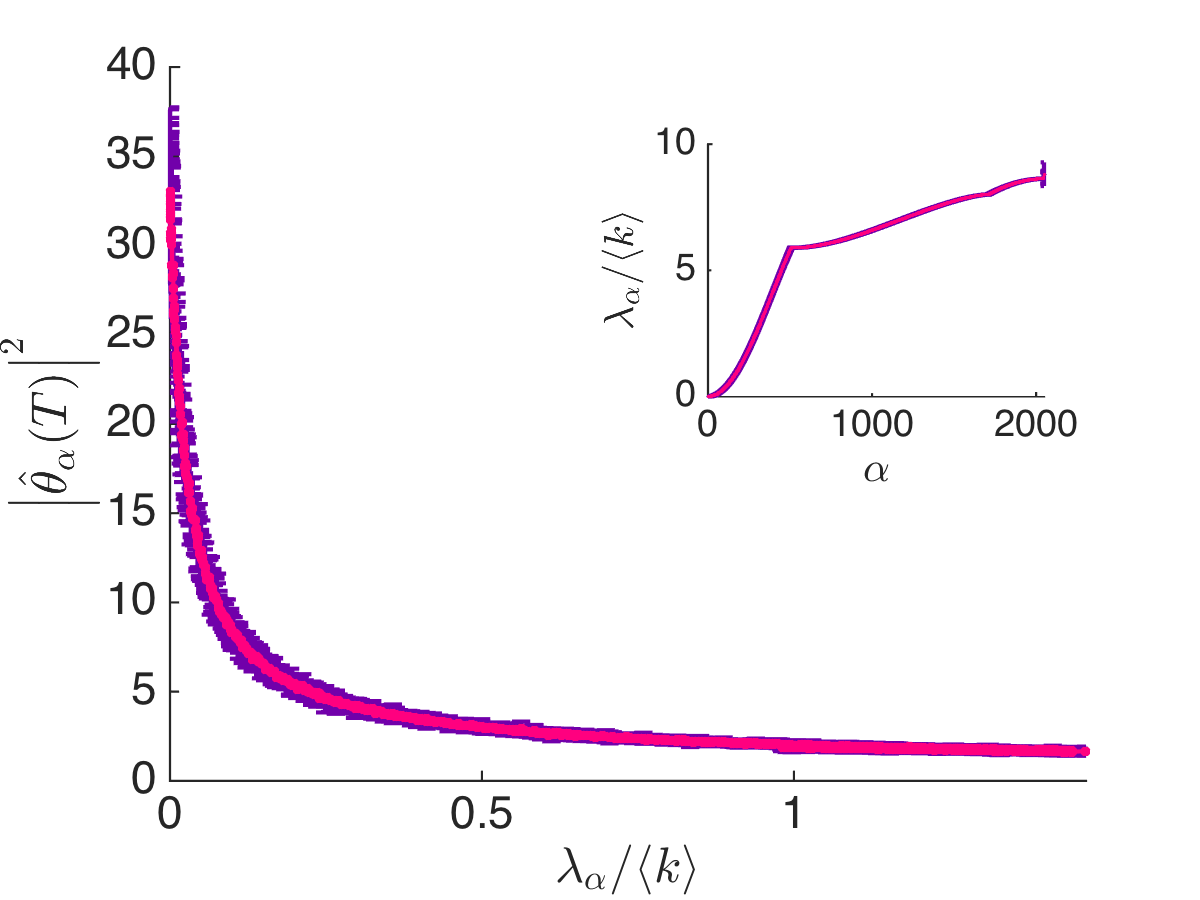}\put(-77,-12){c)}
	\includegraphics[width=0.3\textwidth]{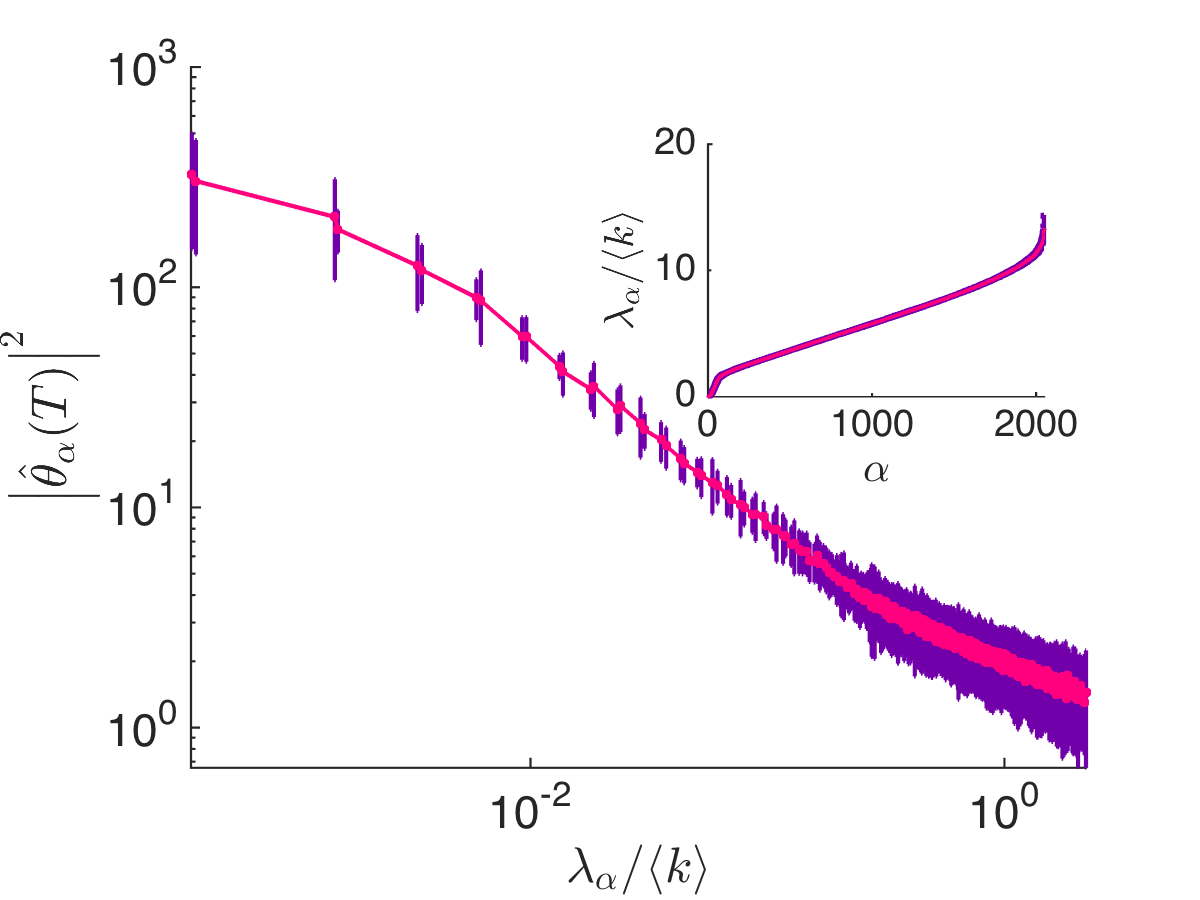}\put(-77,-12){d)}
	\includegraphics[width=0.3\textwidth]{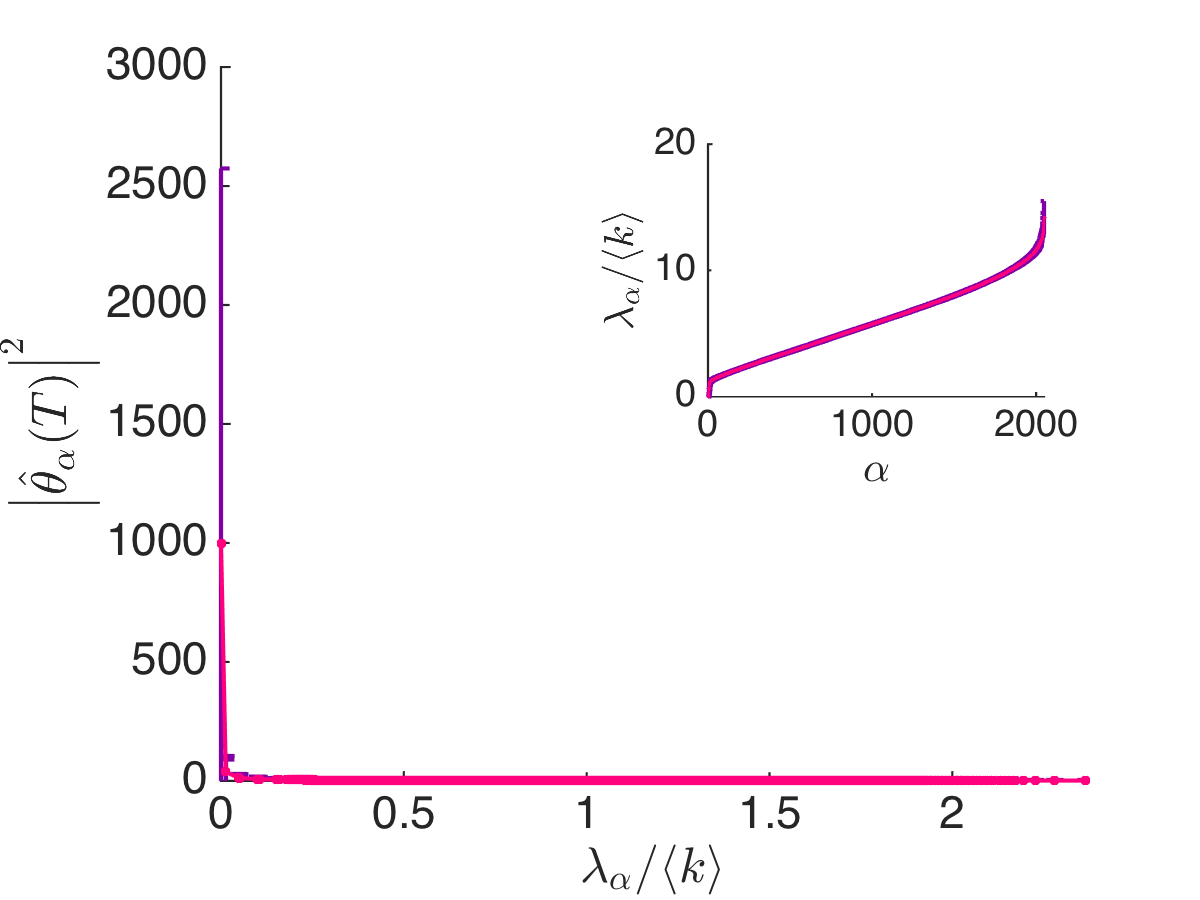}\put(-77,-12){e)}
	\caption{In this figure, we show the average spectra (insets) and average spatial power spectra for $n=10$ realisations of Watts and Strogatz and Lace networks, with error bars in purple. The small magnitude of the variability of the eigenvalues across realisations justifies the averaging of the power spectra across realisations, which are shown in each figure, with error bars. The figures presented here make evident the robustness of our findings with respect to noise that could be introduced by different realisations of the same network model. Top row: Small-World networks (a) non-magnetised, (b) magnetised. Bottom row: Lace networks (c) non-magnetised, (d) supra-oscillating, (e) magnetised}\label{fig:spectra_powa_stat}
\end{figure*}

\begin{figure*}
	\includegraphics[width=0.3\textwidth]{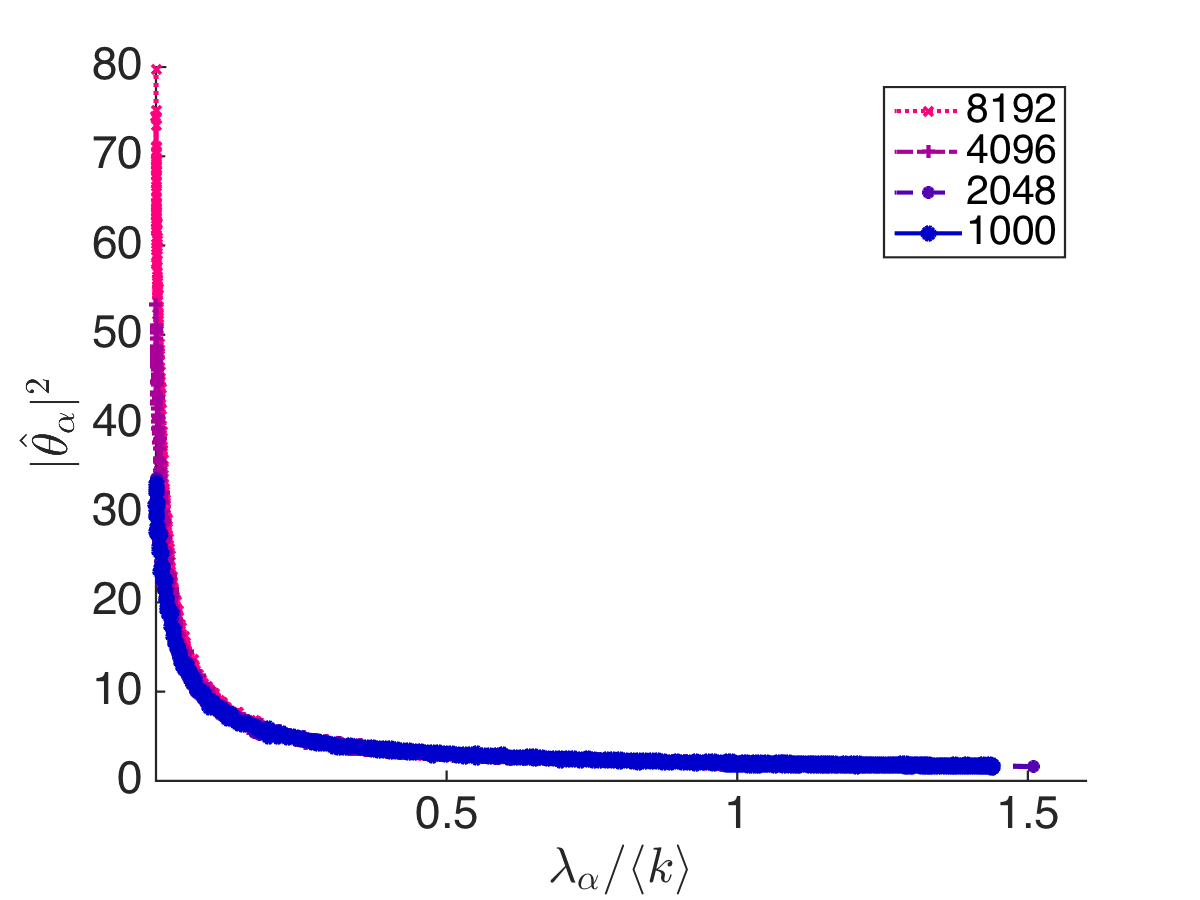}\put(-77,-12){a)}
	\includegraphics[width=0.3\textwidth]{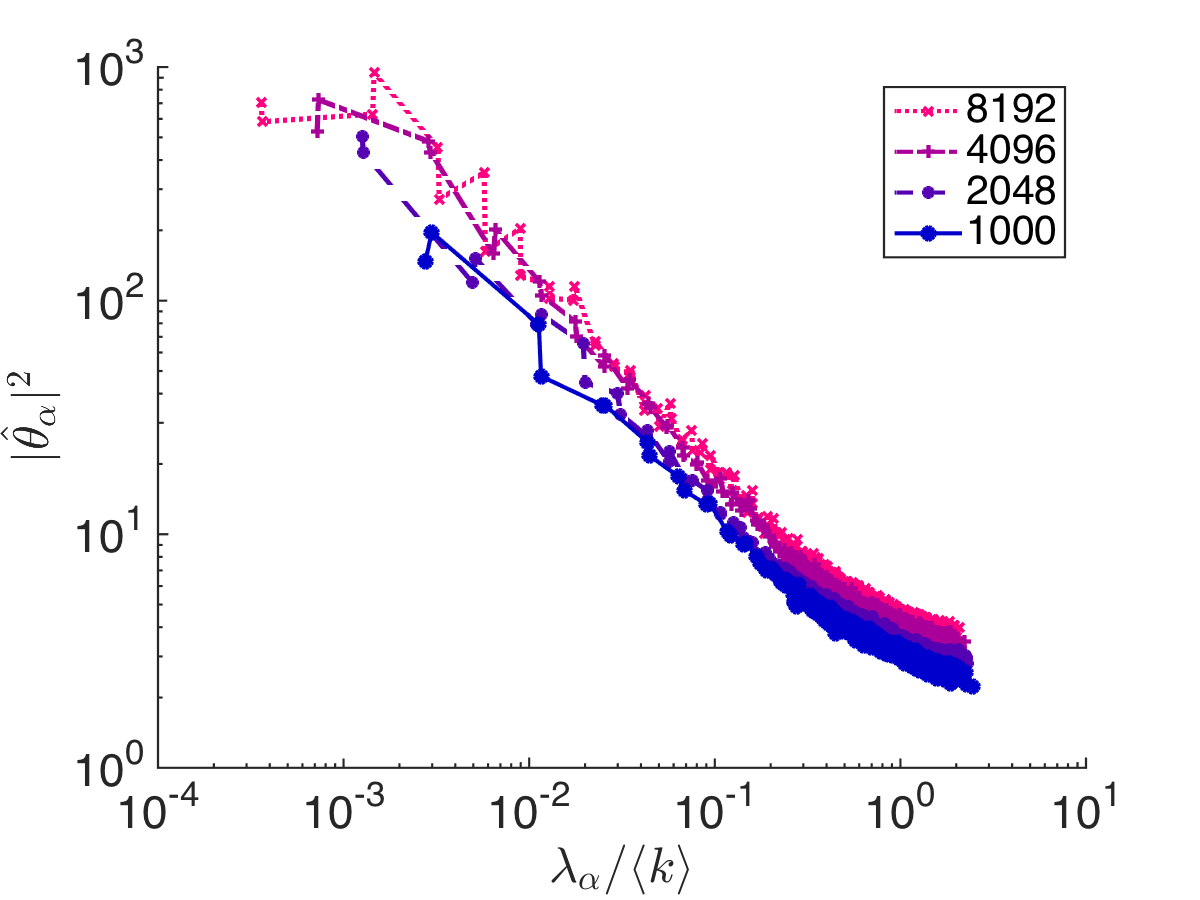}\put(-77,-12){b)}
	\includegraphics[width=0.3\textwidth]{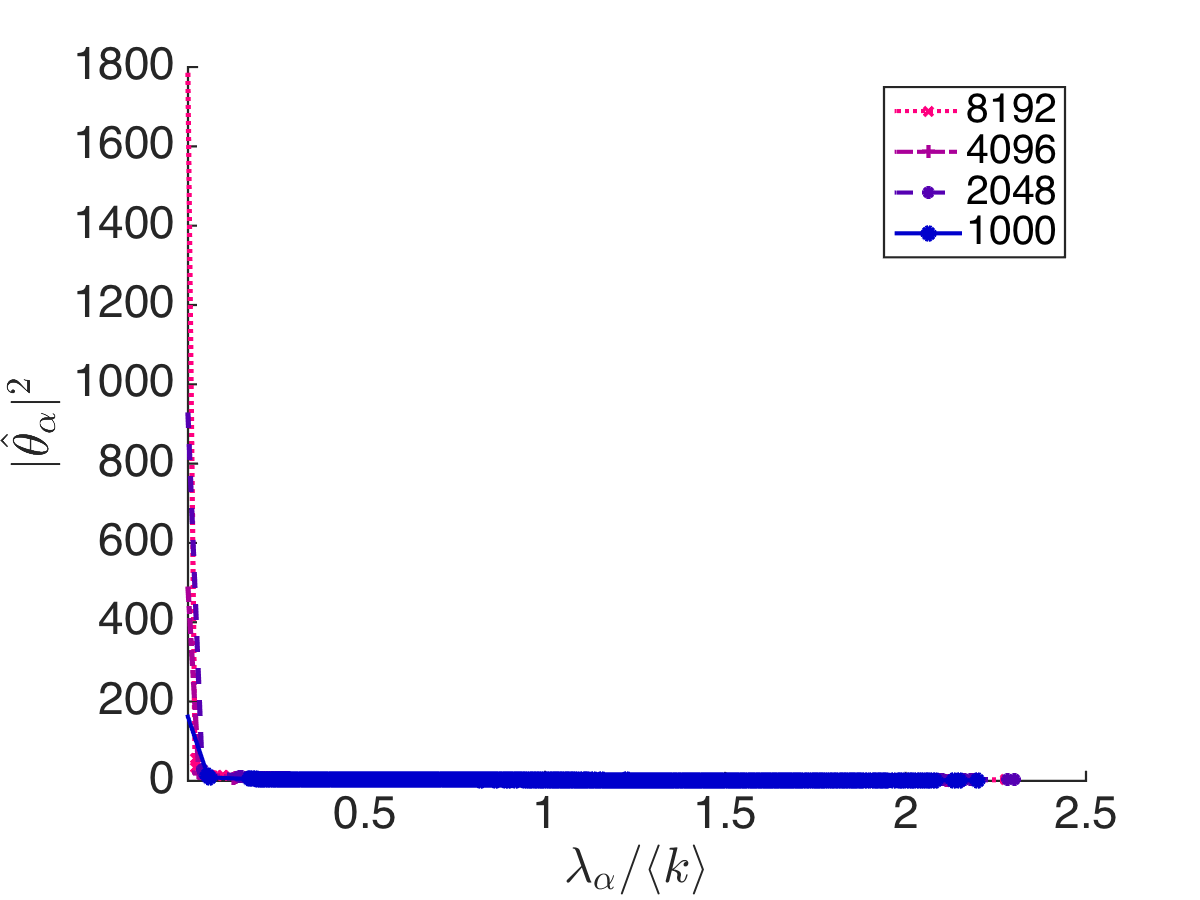}\put(-77,-12){c)}
	\caption{Effect of the system size on the power spectra for the (a) non-magnetised, (b) supra-oscillating, and (c) magnetised states. The underlying are Lace networks of sizes 1000, 2048, 4096 and 8192 nodes with parameters: non-magnetised $\delta=0.5,\ p=0.0001$, supra-oscillating $\delta=0.5,\ p=0.5$ and magnetised $\delta=0.75,\ p=0.5$. We used a log-log scale for (b) to emphasise the particular participation of the modes to the spectra.}\label{fig:powa_spectra_size_lace}
\end{figure*}

The first step of our analysis aims at characterizing the stationary states and their fluctuations: a finite magnetisation triggers the tendency of the spins, on average, to rotate coherently, thus displaying a symmetry breaking, while they rotate almost incoherently when $M \sim 0$. As we are interested in characterising the stationary states of the $XY$-model, we detrend the time series of the spins $\vect{\theta}(t)\in [0,2\pi]^N$ by removing their temporal mean value $\bar{\vect{\theta}}$.
We therefore express the detrended time series using the decomposition in Eq.~\ref{eq:GST_x} and compute the power spectrum $|\hat{\theta}_{\alpha}(t)|^2$, and considered its temporal average
\begin{equation}
\bar{I}_\alpha \equiv \frac{1}{t_{f}-t_{f}/2} \sum_{s=t_{f}/2}^{t_{f}} |\hat{\theta}_{\alpha}(s)|^2,
\end{equation}
which we refer to as the spatial power spectrum. We can thus associate a power spectra to each macroscopic phase. We will now detail the Laplacian spectra and power spectra for each macroscopic state shown in Fig.~\ref{fig:spectra_powa_across}. We observe that the power spectra for each state are strikingly similar for the different topologies, which are nevertheless characterised by different spectra and eigenbasis, as we display in Figs.~\ref{fig:spectra_powa_across}-\ref{fig:spectra_powa_stat}. This strongly suggests the existence of some specific substructures that drive the spatial power spectrum and that must be common to all networks on which the dynamics is run.\\
\textit{Magnetised}\hspace{0.2cm} The Laplacian spectrum for the $k$-regular graph is highly degenerated, reflecting the regularity of the network, while the spectra for the WS and the Lace Networks are very similar, showing similar structure and non-degeneracy due the the random rewiring. On the other hand the power spectra are unsurprisingly largely dominated by the first eigenvalue as it represents the constant component of the signal which dominates as the magnetisation is essentially constant. While the spectra for the Lace and WS networks are very close, the spectra for the $k$-regular network is very different, strengthening our point that specific substructure, potentially independent of the network model considered drive the macroscopic dynamics.\\
\textit{Non-magnetised}\hspace{0.2cm} This state is akin to a random state, as the spins only weakly interact and no long-range order is present. This is directly reflected by the small contributions from all eigenmodes, as there is barely an order of magnitude difference between the largest and smallest amplitudes. The contribution of the eigenvalues decay monotonously and slower than algebraically, and the spectra for the three network models are very similar, the Lace and WS spectra are even quasi overlapping.\\
\textit{Supra-oscillating}\hspace{0.2cm} The power spectrum signature of the state, that only exists for the $k$-regular and Lace network is the most interesting. The fat tail and its seemingly power law decrease hints to a notion of hierarchy in the spatial modes that explain the non tamable oscillating patterns, as the magnetisation is eventually a result of the superposition of all the spatial modes.

It is worth stressing once more that the persistence of the power spectra shape across the $k-$regular, WS and Lace networks is highly non trivial: those networks are fundamentally different from a structural point of view and these differences are mirrored by their dissimilar spectra. Nevertheless, our analysis shows how the stationary dynamics of the $XY$ model selects specific eigenvectors, whose properties are likely shared by these graphs.

To conclude this section, it is interesting to note that although the contribution of the eignenmodes decreases with the eigenvalues, they do so non-monotonically. We investigate this phenomena in the next section.

\subsection{Consistency across network realisations}\label{subsec:consistency}
The WS and Lace networks have an element of randomness in their construction. It is therefore crucial to verify that the properties of the spectra and power spectra we observed in the previous section are not accidental, but genuinely representative of a class of networks. We generated $n=10$ realisations of the two types of networks in each state they can support, see Fig.~\ref{fig:spectra_powa_stat}. The spectra of the Laplacians are remarkably consistent, as shown by the small error bars. This small magnitude of the variability of the eigenvalues across realisations justifies the averaging of the power spectra across realisations. It is remarkable that this variance, affecting particular structures of the networks, does not have any effect on the power spectra, as they are all very consistent with low variance, except for some noise at the beginning of the power spectra. The emerging macroscopic properties are not affected by the local differences induced by the variance and the structural differences between the Lace and WS, that make Lace networks support the supra-oscillating state and not the WS, are robust to the noise that is introduced by different realisations of the same network model.

Earlier, we pointed out the non-monotonic decrease of the eigenmodes amplitude with the eigenvalues, on top of a clear overall decreasing trend. On the one hand, these fluctuations could be due to stochastic effects of particular network realisation which are ironed out when an ensemble average is taken. On the other hand, they could be genuine and due solely to the dynamics. To investigate the cause of these fluctuations, we averaged the power spectra for the different realisation of Lace networks, and observe that both scenarios happen. In the case of the magnetised and supra-oscillating states, the curves becomes very smooth and decreases monotonically, and the non-magnetised power spectrum remains intrinsically noisy. This is not particularly surprising, as the first two states contain some degree of order and even a handful of realisations are enough to even out the fluctuations. On the contrary, the behaviour of the spins in the non-magnetised state is essentially uncorrelated. This randomness is heightened by the randomness inherent to the generation of the Lace networks, and the power spectrum strongly carries the mark of this structural randomness, contrary to the case of the two other states, where the temporal structure, induced by underlying network structure, is enough to cancel the variations in the structure.
Finally, in Fig.~\ref{fig:powa_spectra_size_lace}, we present evidence that the power spectral signatures for the Lace networks are not due to finite size effects. The shape of the power spectra and the relative importance of the eigenmodes are consistent for networks of sizes $N=1000, 2048, 4096, 8192$.

\section{Conclusion}\label{sec:conclusion}
In this paper, we presented the temporal Graph Signal Transform (tGST), a method to decompose time dependent signals living on the nodes of a network, using a basis that incorporate structural information. We applied tGST to the time series of the spins of the $XY$ spin model in its three possible macroscopic states on three different network topologies. We found clear spatial power spectral signatures that characterise each state. Importantly, these signatures are robust across topologies and to structural variability in different realisations of the Watts-Strogatz and Lace networks. In all cases, the power spectra are dominated by small eigenvalues, that correspond to smoother structures. The shape of the power spectra and their decrease reflect the behaviour of the macroscopic magnetisation of the three states in Fig.~\ref{fig:macro-states}: the only significant contribution of the magnetised state is the constant eigenvector; the non-magnetised state is also dominated the constant eigenvector, but there are non negligible contribution from higher modes, whose power decays exponentially. This is consistent with the notion that in the non-magnetised state, the spins oscillate in a random fashion. Finally, the power spectrum of the supra-oscillating state displays a power law like decay, hinting that a hierarchy of modes exists and elucidating the origin of this state.

These results offer a new avenue to characterise not only macroscopic states in statistical mechanics models but also the behaviour of real world system. This technique is powerful enough to circumvent traditional problems such as the need to use finite size scaling to take into account finite size effects. This study constitutes the first step to quantify and identify key network features that support collective states and opens many questions to fully understand this new tool. We are now investigating the characterisation of the structures of the eigenvectors to clearly pinpoint the key mesoscopic structures that supports the dynamics, effectively constituting a centrality measure for network structures. A parallel line of investigation is the combination of spatial and temporal frequencies to define dispersion relations for networks, potentially giving a simple criterion to classify networks. Finally, we plan to investigate the effect of the basis chosen for the decomposition of the signal, as different basis will emphasis different properties of the original signal.

\subsection*{Acknowledgements} P.E. acknowledges financial support from a PET methodology programme grant from MRCUK (ref no. G1100809/1). P.E. and T.T. acknowledge financial support for reciprocal visits from a DAIWA Small Grant from the DAIWA foundation. This work was partly supported by Bilateral Joint Research Project between JSPS, Japan, and F.R.S.–FNRS, Belgium.

\end{document}